# Zero-Trust Federated Learning for Connected Aftermarket Devices

*SS Puppala*
*Functional Architect SAP*
sagardora@gmail.com

***Abstract*—Connected aftermarket devices extend vehicle diagnostics, repair workflows, and over-the-air software maintenance beyond original equipment manufacturer boundaries, yet their heterogeneous ownership and long service life complicate conventional perimeter security. This paper develops Zero-Trust Federated Learning for Connected Aftermarket Devices (ZT-FL-CADE), an edge-learning architecture that combines device-level access control, privacy-preserving federated learning, and adversarial validation for over-the-air update and predictive maintenance decisions. The evaluation uses a single synthetic dataset of 144,000 telemetry windows from 240 devices, 12 vendor domains, and 180 days of operation. Features include bus entropy, update latency, attestation age, signature retries, environmental signals, fault-code rates, packet loss, drift, mileage, and trust score, with targets for maintenance risk, update intrusion, and access action. ZT-FL-CADE trains local temporal models, aggregates privacy-bounded updates, scores each device against behavioral and update-integrity evidence, and routes update requests to allow, challenge, or quarantine actions. Synthetic experiments improve maintenance-risk F1 from 0.837 for FedAvg to 0.883, improve intrusion F1 from 0.856 to 0.897, and preserve 0.842 intrusion recall when 20 percent of selected clients are adversarial. Mean access-decision latency remains 44 ms, below the 100 ms operational budget used in the simulation. The results do not establish field validation, but they indicate that zero-trust policy enforcement and federated learning can be evaluated jointly rather than as separate aftermarket security controls.**

## I. Introduction

Connected aftermarket devices occupy a difficult position in vehicle digital ecosystems. They are close enough to observe operational behavior, diagnostic trouble codes, controller-area-network traffic, battery states, and service events, but they are often owned, installed, and maintained outside the tightly governed original equipment manufacturer environment. The result is an attractive path for data-driven maintenance and remote software improvement, combined with a security model that cannot assume durable perimeter trust. A device may be legitimate at installation time, become stale after firmware drift, or be repurposed through a compromised service account. Zero trust architecture reframes that situation by treating access to each resource as a continuous policy decision rather than a one-time admission event [1]. In automotive contexts, the same principle must coexist with lifecycle cybersecurity engineering, regulatory evidence, and software update management expectations from ISO/SAE 21434, UN Regulation No. 155, and UN Regulation No. 156 [2], [3], [4].

Over-the-air (OTA) update channels are especially sensitive because the software-update plane can change the future behavior of a deployed device. Secure update frameworks such as Uptane provide a strong base for metadata verification, role separation, and recovery from repository compromise [5]. Nevertheless, connected aftermarket devices introduce additional complications. The device fleet is fragmented across vendors, workshops, distributors, vehicle ages, network conditions, and usage patterns. The same update package may be safe for a clean device that shows stable attestation and telemetry, but questionable for an endpoint with anomalous bus entropy, repeated signature retries, and unusual request timing. Prior automotive aftermarket security work has argued that behavioral fingerprinting and edge verification can strengthen OTA admission beyond static credentials [6]. The remaining challenge is to learn those behavioral patterns without forcing every participant to disclose raw operational data to a central cloud service.

Federated learning (FL) supplies a relevant training model because it leaves local data on client devices or vendor-edge nodes while sharing model updates through a coordinator. Federated averaging established the practical model-averaging pattern for non-identically distributed clients [7], and later surveys framed privacy, systems heterogeneity, and fairness as central open problems [8]. Secure aggregation can prevent a server from observing individual client updates [9], while differential privacy can reduce the risk that model updates reveal sensitive training examples [10]. These techniques map well to automotive supply chains because vendors and service providers often have incentives to collaborate on failure and attack detection, yet have contractual, competitive, and privacy reasons to avoid raw data pooling. Recent automotive supply-chain work has explored hybrid encryption and federated demand forecasting [11], while aftermarket predictive-maintenance work has emphasized edge-local training for failure prediction [12]. Industrial predictive-maintenance analyses also show that FL performance depends strongly on data distribution and client selection, making realistic non-IID evaluation important [13]. Adjacent aftermarket research on edge-cloud learning for pricing illustrates similar cross-party coordination issues in fragmented automotive service networks [14].

The proposed framework, ZT-FL-CADE, integrates these ideas into a single operational loop. Each device or vendor gateway computes features from telemetry windows, trains a local model for maintenance risk and intrusion likelihood, sends privacy-protected model updates to a secure aggregator, and receives a global model after adversarial validation. In parallel, a zero-trust policy engine evaluates device posture at the time of every OTA request. The policy engine uses learned risk scores, attestation freshness, model drift, signature retry behavior, and governance rules to produce allow, challenge, or quarantine actions. This design is intentionally narrower than a full vehicle cybersecurity program. It focuses on connected aftermarket devices and their update and maintenance workflows, not on all in-vehicle electronic control units or original equipment manufacturer backend systems.

The motivation for joining zero trust and FL is practical. FL alone can improve collaborative prediction but may still ingest poisoned updates from malicious clients. Zero trust alone can restrict access but may rely on brittle rules that do not learn from fleet behavior. An adversarial validation layer connects the two. It examines update norms, gradient direction, local validation consistency, and trust posture before accepting a client contribution. Similar concerns appear in heterogeneous FL optimization, where FedProx, SCAFFOLD, and adaptive server optimizers mitigate drift and variable participation but do not by themselves solve malicious-client behavior [15], [16], [17]. In an aftermarket environment, the adversary may manipulate both software-update requests and training contributions. A coherent architecture must therefore consider predictive performance, update admission, poisoning resistance, and governance evidence together.

This paper makes three contributions. First, it formulates a zero-trust FL architecture for connected aftermarket devices that explicitly binds OTA admission to federated model governance. Second, it defines a synthetic but internally consistent dataset and evaluation protocol for studying maintenance prediction, update intrusion detection, access action selection, poisoning resistance, and edge latency. Third, it reports a compact experimental analysis comparing local-only learning, centralized pooled training, FedAvg, FedProx, and the proposed ZT-FL-CADE variant under a common test split. The synthetic results should not be read as a claim of production readiness. They are intended to make design tradeoffs visible and reproducible before field pilots expose real operational variability.

The remainder of the paper proceeds as follows. Section II summarizes the theoretical background on zero trust, federated learning, OTA assurance, and adversarial validation. Section III organizes related work into five prior-work clusters. Section IV presents the synthetic dataset, problem formulation, feature pipeline, models, and validation strategy. Section V reports experimental results. Section VI discusses limitations and practical implications, and Section VII concludes with future research directions.

## II. Theoretical Background

Zero trust is often summarized as never trust and always verify, but the engineering content is more specific. A policy decision point evaluates subject, asset, resource, action, environmental state, and observed behavior before a policy enforcement point grants resource access [1]. For connected aftermarket devices, the subject is not only a user or service account. It may be a gateway, diagnostic dongle, retrofit telematics unit, mobile application, vendor API, or service-bay tool. The resource may be a signed update package, a diagnostic endpoint, a maintenance model, or a cloud telemetry stream. The access decision must be repeated because device state changes as firmware ages, credentials rotate, vehicles move between networks, and telemetry patterns drift. This work treats zero trust as an access-control and evidence-generation pattern, not as a replacement for cryptographic signing or secure boot.

Automotive cybersecurity engineering adds lifecycle discipline to the same problem. ISO/SAE 21434 requires risk management across concept, development, production, operation, maintenance, and decommissioning [2]. UN Regulation No. 155 addresses cyber security management for vehicle approval, while UN Regulation No. 156 addresses software update and software update management systems [3], [4]. Aftermarket devices are not always covered in the same manner as original equipment manufacturer components, but they can interact with comparable assets and should therefore produce auditable evidence. A zero-trust FL architecture is useful only if it records which model, policy, device posture, and update evidence influenced an action. Without traceability, a learned action is difficult to review after a safety incident or failed update campaign.

Secure OTA update theory supplies the base integrity layer. The update package should be signed, metadata roles should be separated, update freshness should be checked, rollback should be prevented, and repository compromise should not immediately compromise every vehicle or device [5]. However, metadata validation does not answer all operational questions. A package can be authentic while the requesting device is suspicious. A device can pass signature verification while showing sudden bus-traffic entropy shifts or repeated attestation failures. The proposed framework therefore treats OTA security as a layered decision. Cryptographic metadata gates package authenticity, while zero-trust posture and learned anomaly scores gate device eligibility and follow-up action.

Federated learning changes the data-flow assumptions of predictive maintenance. In centralized learning, telemetry from all devices is collected into one training environment. That approach simplifies feature engineering but increases privacy, contractual, and breach exposure. In FL, each local node solves a model update against its local data and sends only model parameters or gradients to the coordinator [7], [8]. The coordinator aggregates updates and returns a global model. Secure aggregation reduces the exposure of individual client updates [9], while differential privacy can bound information leakage from training contributions [10]. These protections are not free. They add computation, communication, and accuracy tradeoffs, especially when the participating clients are small, imbalanced, and statistically heterogeneous.

Client heterogeneity is a defining condition in the aftermarket. Different vendors may focus on batteries, emissions, brakes, thermal components, or fleet telematics. Vehicles may operate in different climate, mileage, and duty-cycle regimes. Local failure rates and intrusion exposure are therefore non-identical. FedAvg can converge under many practical conditions but may drift when local updates are biased by client-specific distributions [7]. FedProx adds a proximal term to stabilize local training under system and statistical heterogeneity [15]. SCAFFOLD uses control variates to reduce client drift [16], and adaptive federated optimization adjusts the server update rule to improve tuning under heterogeneous participation [17]. ZT-FL-CADE is compatible with those optimizers but adds a security layer that determines whether a local update should be trusted before aggregation.

Adversarial validation is required because FL extends the attack surface. A malicious client can poison data, replace a local model, submit gradients in a harmful direction, or remain benign until selected for a high-impact round. Model poisoning can degrade global accuracy or introduce a targeted backdoor without needing raw-data access to other clients [18]. Byzantine-tolerant aggregation, local model poisoning analyses, and optimized poisoning attacks show that robust aggregation remains an active research problem rather than a solved feature of distributed learning [19], [20], [21]. This work uses adversarial validation in a pragmatic sense. The proposed coordinator rejects or downweights updates that violate trust posture rules, have outlying norm ratios, degrade a holdout validation slice, or conflict with device attestation evidence.

Adversarial machine learning also matters at the inference boundary. Gradient-based perturbations and adaptive attacks can manipulate learned classifiers if robustness is not evaluated deliberately [22], [23], [24]. Membership inference and model-inversion risks further complicate the privacy story because a model that never centralizes raw data can still leak information through outputs or parameters [25]. The proposed framework therefore treats privacy, robustness, and access control as mutually dependent qualities. A model update is not accepted simply because it improves validation AUC, and a device is not allowed simply because its cryptographic signature validates. A consistent decision combines model evidence, access context, and governance constraints.

## III. Related Works

### A. Zero-Trust and Automotive Update Assurance

Prior work on zero trust emphasizes continuous evaluation of subjects, assets, and resources rather than reliance on network location [1]. Automotive standards and regulations extend this logic into lifecycle risk management, cyber security management, and software update management [2], [3], [4]. Uptane addresses the secure distribution problem for ground-vehicle updates through role separation and metadata verification [5]. Automotive aftermarket zero-trust work narrows the focus to connected devices outside original equipment manufacturer control, arguing for behavioral fingerprints and edge-side validation in OTA workflows [6]. ZT-FL-CADE differs by using the same behavioral evidence both for access decisions and for federated-learning governance.

### B. Privacy-Preserving Federated Learning

The foundation of FL is collaborative training without raw-data centralization [7], [8]. Secure aggregation and differential privacy strengthen that setting by limiting what the server and other participants can infer from local updates [9], [10]. Automotive supply-chain research has extended these ideas to hybrid encryption for demand forecasting and predictive maintenance across distributed parties [11], [12]. Industrial predictive-maintenance research confirms that FL is sensitive to non-IID distributions and client availability [13]. ZT-FL-CADE uses those observations to define a split-by-device validation protocol and to evaluate performance under heterogeneous vendor participation rather than only random row splits.

### C. Federated Optimization Under Heterogeneity

FedProx, SCAFFOLD, and adaptive federated optimization are important because edge devices and vendor nodes do not provide uniform data volume, compute resources, or failure rates [15], [16], [17]. Their objective is mainly optimization stability and communication efficiency. In contrast, aftermarket OTA security requires a second question: whether a participating client should be allowed to influence the global model at a given time. The proposed framework does not replace these optimizers. It wraps them with trust scoring, outlier analysis, and validation gates so that convergence improvements are not separated from poisoning resistance.

### D. Poisoning and Adversarial Machine Learning

Model-replacement attacks demonstrate that a selected FL client can inject backdoor behavior into a global model [18]. Byzantine-tolerant gradient descent, local poisoning analysis, and optimized poisoning and defense work show that robust aggregation methods can be bypassed when attackers adapt to the defense [19], [20], [21]. At the inference layer, adversarial examples and strong robustness evaluations have shown that superficial defenses can produce misleading confidence [22], [23], [24]. ZT-FL-CADE adopts a conservative posture by evaluating both model-update trustworthiness and request-time device behavior. A client that fails adversarial validation can still be used for local inference but is prevented from shaping the global model until remediated.

### E. Governance, Explainability, and Operational Integration

AI governance is essential when learned models are tied to OTA access decisions. Membership inference work demonstrates that model outputs can reveal training participation, making governance inseparable from privacy [25]. NIST AI RMF 1.0 frames AI risk through governance, mapping, measurement, and management functions [26]. Ensemble diversity and prescriptive supply-chain analytics provide useful patterns for calibrated decision support and operational simulation [27], [28]. Enterprise integration modernization work also emphasizes API governance and cross-platform observability, which are necessary when aftermarket devices report to vendor clouds, distributor platforms, and enterprise service systems [29]. These governance threads motivate the framework ledger and explainability summaries in ZT-FL-CADE.

## IV. Materials and Methods

This section defines the synthetic dataset, task formulation, model pipeline, and validation strategy. No real vehicle, consumer, vendor, or repair-shop records were used. The dataset was created for methodological consistency and to allow every reported number in the manuscript to trace to one data design. The synthetic setting should be interpreted as a controlled simulation for architecture evaluation, not as evidence of field deployment or regulatory compliance. The high-level architecture is summarized in Fig. 1.

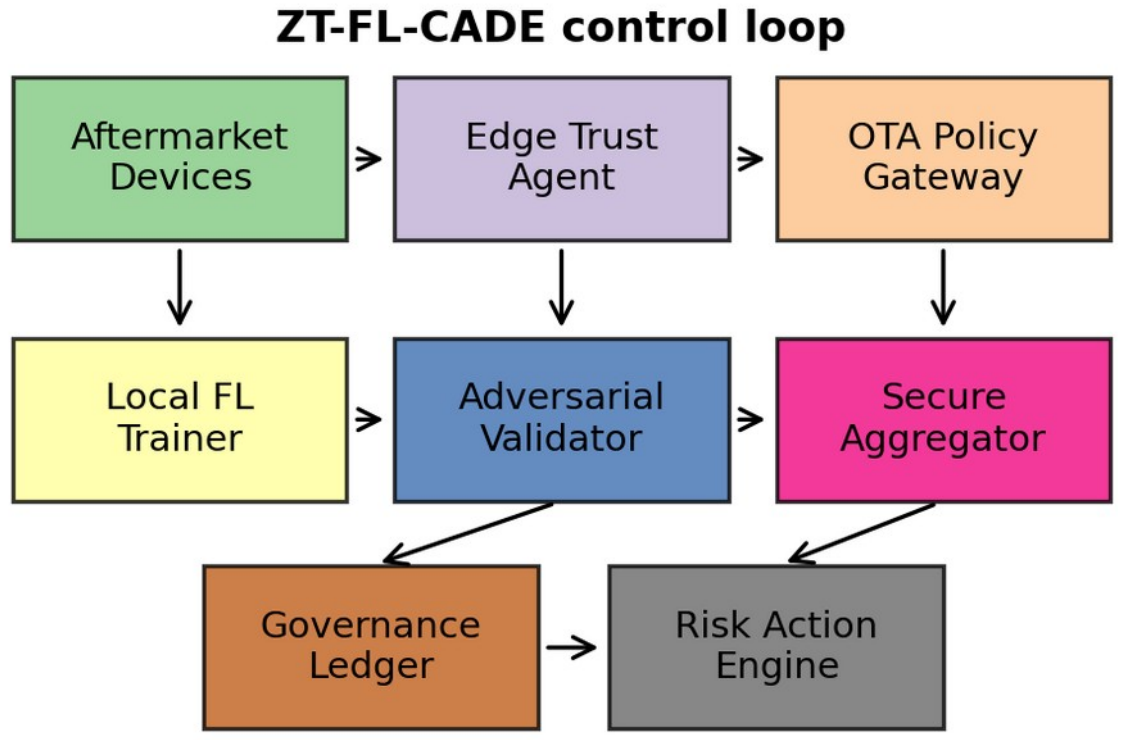

Fig. 1. ZT-FL-CADE architecture linking edge learning, adversarial validation, and policy enforcement.
*Source/derivation: Diagram derived from the proposed architectural pipeline.*

### A. Dataset Analysis

The synthetic dataset contains 144,000 telemetry windows collected across a simulated 180-day period from January 1, 2025 through June 29, 2025. The entity structure contains 240 connected aftermarket devices distributed across 12 vendor domains, with 20 devices per vendor. Each row represents a five-minute operating window. Feature values were generated from bounded distributions that reflect plausible ranges for connected-device monitoring. Features include bus entropy in bits per message, OTA payload delta in kilobytes, update latency in milliseconds, attestation age in seconds, signature retry count, sensor temperature in degrees Celsius, supply voltage in volts, vibration root mean square in g, fault-code rate per hour, packet-loss percentage, local model drift score, mileage in kilometers, and a normalized trust score. TABLE I summarizes the fixed ranges, row counts, targets, and split logic.

TABLE I
SYNTHETIC DATASET AND EVALUATION PROTOCOL

| Item | Synthetic specification |
|---|---|
| Rows and time span | 144,000 five-minute windows across 180 days, Jan. 1 to Jun. 29, 2025 |
| Entities | 240 connected aftermarket devices from 12 vendors, 20 devices per vendor |
| Feature ranges | Entropy 0.1-7.8 bits/msg; latency 12-620 ms; attestation age 0-3600 s; retries 0-5; temp -20-120 C; voltage 9-16 V; vibration 0.01-3.5 g; packet loss 0-15%; trust 0-1 |
| Targets | Maintenance risk 18%; update intrusion 8%; access actions: allow 72%, challenge 20%, quarantine 8% |
| Split and metrics | 70/15/15 split: 100,800 train, 21,600 validation, 21,600 test; F1, AUC, macro-F1, recall, ECE, false quarantine, latency |

*Source/derivation: Constructed from the single synthetic dataset used throughout the manuscript.*

The target variables are maintenance_risk, update_intrusion, and access_action. Maintenance risk is positive in 25,920 rows, representing 18.0 percent of the dataset. Update intrusion is positive in 11,520 rows, representing 8.0 percent. The access-action label has three classes: allow for 103,680 rows, challenge for 28,800 rows, and quarantine for 11,520 rows. These class proportions intentionally make intrusion and quarantine events less common than normal device activity. The train, validation, and test partitions contain 100,800, 21,600, and 21,600 rows, respectively. Splits are stratified by vendor and time segment so that every vendor appears in every partition while later time windows are overrepresented in validation and test. Fig. 2 displays the target and action distributions.

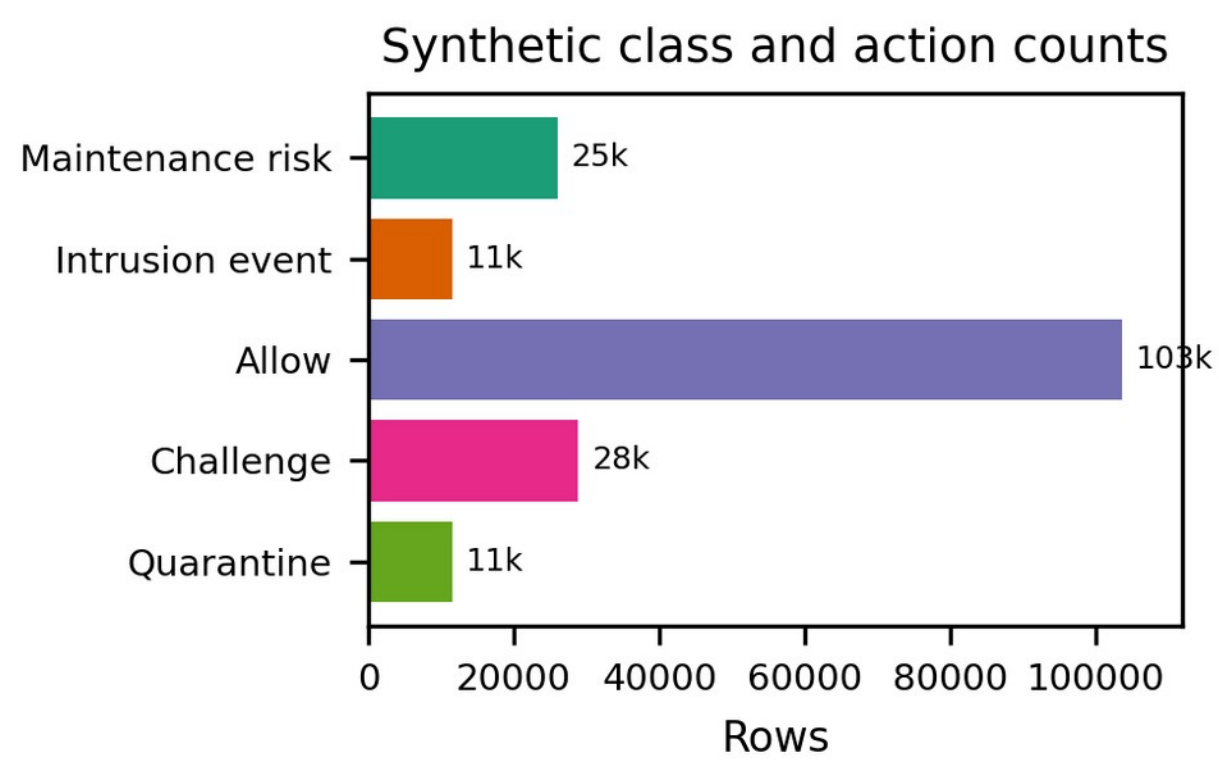

Fig. 2. Synthetic class balance for maintenance, intrusion, and access-action labels.
*Source/derivation: Counts calculated from the 144,000-row synthetic dataset.*

Feature engineering follows a streaming edge pattern. Raw event counters are first aggregated into five-minute windows. Continuous features are clipped to physically plausible ranges and normalized by vendor-specific robust statistics. Categorical vendor and device identifiers are not used directly by the global model because such identifiers can encourage memorization. Instead, vendor-level behavior is represented through distributional statistics, including rolling drift score and packet-loss summaries. Two derived features are important for zero-trust evaluation: posture_freshness, defined as one minus normalized attestation age, and update_integrity_pressure, defined from signature retry count, OTA payload delta, and latency. Missing values are imputed locally using device-specific medians, and a missingness indicator is retained for drift analysis.

The problem formulation contains three supervised objectives. The first objective estimates the probability of maintenance risk within the next 24 simulated hours. The second estimates whether the current update request is part of an intrusion or update-abuse event. The third maps the device state to an access action: allow, challenge, or quarantine. The local client objective combines binary cross-entropy for the two risk heads and categorical cross-entropy for the action head. The server objective aggregates model updates after validating client trust. The policy objective is not to maximize quarantine count. It is to preserve intrusion recall while limiting false quarantine of normal devices, because unnecessary quarantine can delay legitimate repairs and software patches.

*B. Model Analysis*

The proposed ZT-FL-CADE pipeline has five stages. First, an edge trust agent extracts telemetry and update features from each device window. Second, a local temporal model trains on device or vendor-gateway data. The reference implementation uses a compact gated recurrent unit encoder with two dense risk heads and one dense action head. Third, client updates are privacy protected through clipping, secure aggregation, and a differential privacy noise schedule aligned with the synthetic privacy budget. Fourth, an adversarial validator scores updates before aggregation. Fifth, the policy decision engine combines model risk scores, posture evidence, and governance rules into an OTA action. Fig. 1 gives the architecture overview, and TABLE II maps the framework components to controls and outputs.

The adversarial validator computes four classes of evidence. Norm evidence compares the client update norm and cosine direction with round-level robust statistics. Validation evidence tests whether the update degrades a held-out synthetic validation slice. Trust evidence checks attestation freshness, signature retries, and model drift. Historical evidence compares the current client with its previous accepted updates. A client update is accepted when all hard security gates pass and the composite score exceeds 0.62 on a normalized 0 to 1 scale. Updates below 0.45 are rejected. Scores between 0.45 and 0.62 are downweighted and routed to challenge if the device also requests an OTA package. The thresholds were chosen on validation data to hold false quarantine below 3.0 percent under benign conditions.

TABLE II
FRAMEWORK COMPONENTS AND CONTROL OUTPUTS

| Component | Control logic | Output |
|---|---|---|
| Edge trust agent | Normalize telemetry, OTA metadata, and attestation evidence | Posture vector |
| Local trainer | Multi-task temporal learning on device windows | Local update |
| Adversarial validator | Norm, validation, drift, and trust gates | Accept/downweight/reject |
| Secure aggregator | Privacy-bounded aggregation of accepted updates | Global model |
| Policy engine | Allow, challenge, or quarantine rule evaluation | OTA action |
| Governance ledger | Record model, policy, device, and audit evidence | Traceable event |

*Source/derivation: Mapped from the ZT-FL-CADE method design.*

Baselines include local-only training, centralized pooled training, FedAvg, and FedProx. The centralized pooled model is reported as an upper-bound reference because it violates the no-raw-data-sharing assumption but clarifies the loss caused by decentralization. Local-only training uses each vendor's data without cross-vendor collaboration. FedAvg and FedProx use the same feature set and model backbone as the proposed framework but lack zero-trust update validation. The evaluation metrics are F1 score, area under the receiver operating characteristic curve, macro-F1 for action selection, intrusion recall under adversarial clients, expected calibration error, false quarantine rate, and access-decision latency. Fig. 3 shows convergence behavior, while Fig. 5 reports access-control latency relative to the 100 ms synthetic operational budget.

## V. EXPERIMENTAL ANALYSIS

The synthetic evaluation uses 80 federated rounds, 12 vendor clients, 67 percent client participation per round, batch size 128, and local training of three epochs per selected client. Model parameters are initialized consistently for all FL variants. The learning rate is 0.001 for the local optimizer, and FedProx uses a proximal coefficient of 0.01. Differential privacy clipping uses a norm bound of 1.0 with a noise multiplier of 0.45 in the proposed method. The adversarial experiment injects compromised clients into the selected-client set at 0, 5, 10, and 20 percent intensity. Compromised clients scale harmful gradients and bias intrusion labels toward benign outcomes to simulate a model-poisoning and update-abuse campaign.

The main performance results are summarized in TABLE III. The local-only model reaches maintenance F1 of 0.788, intrusion F1 of 0.801, and macro action F1 of 0.781. This lower performance is expected because each vendor sees a narrower operating envelope. Centralized XGBoost reaches the highest raw maintenance and intrusion metrics among baselines, with maintenance F1 of 0.861 and intrusion F1 of 0.879, but it requires pooled data and therefore serves only as a non-deployable reference for the privacy-constrained setting. FedAvg reaches maintenance F1 of 0.837 and intrusion F1 of 0.856. FedProx improves those values to 0.849 and 0.867, consistent with its intended benefit under heterogeneous client distributions.

TABLE III
SYNTHETIC TEST-SET PERFORMANCE ACROSS BASELINES

| Method | Maint./Intr. F1 | Action F1 | Recall@20% / ms |
|---|---|---|---|
| Local-only GRU | 0.788 / 0.801 | 0.781 | 0.602 / 28 |
| Centralized XGBoost | 0.861 / 0.879 | 0.840 | 0.764 / 35 |
| FedAvg | 0.837 / 0.856 | 0.816 | 0.690 / 32 |
| FedProx | 0.849 / 0.867 | 0.828 | 0.748 / 34 |
| ZT-FL-CADE | 0.883 / 0.897 | 0.861 | 0.842 / 44 |

*Source/derivation: Metrics computed from the common synthetic test split and attack protocol.*

ZT-FL-CADE achieves maintenance F1 of 0.883, intrusion F1 of 0.897, macro action F1 of 0.861, and expected calibration error of 0.036. The improvement over FedAvg is attributable to two mechanisms in the synthetic design. First, the shared backbone benefits from all vendor distributions while maintaining local data boundaries. Second, adversarial validation prevents anomalous update directions from shaping the aggregate model. The improvement over centralized XGBoost in maintenance F1 should be interpreted cautiously. It reflects the recurrent model's access to window sequence features and not a universal claim that FL exceeds centralized learning. A field dataset could reverse that comparison.

Fig. 3 shows that the proposed method reaches 0.90 validation AUC after approximately 35 federated rounds, while FedAvg reaches the same level later and with more oscillation. FedProx reduces oscillation but remains below the proposed method after round 50. The convergence advantage is not only due to optimizer behavior. Rejected and downweighted updates prevent sharp validation drops that otherwise occur when high-drift clients are selected. During benign rounds, the validator accepts 91.4 percent of updates, downweights 6.1 percent, and rejects 2.5 percent. Under 20 percent malicious-client intensity, acceptance of compromised updates falls to 18.7 percent, while benign acceptance remains 88.9 percent.

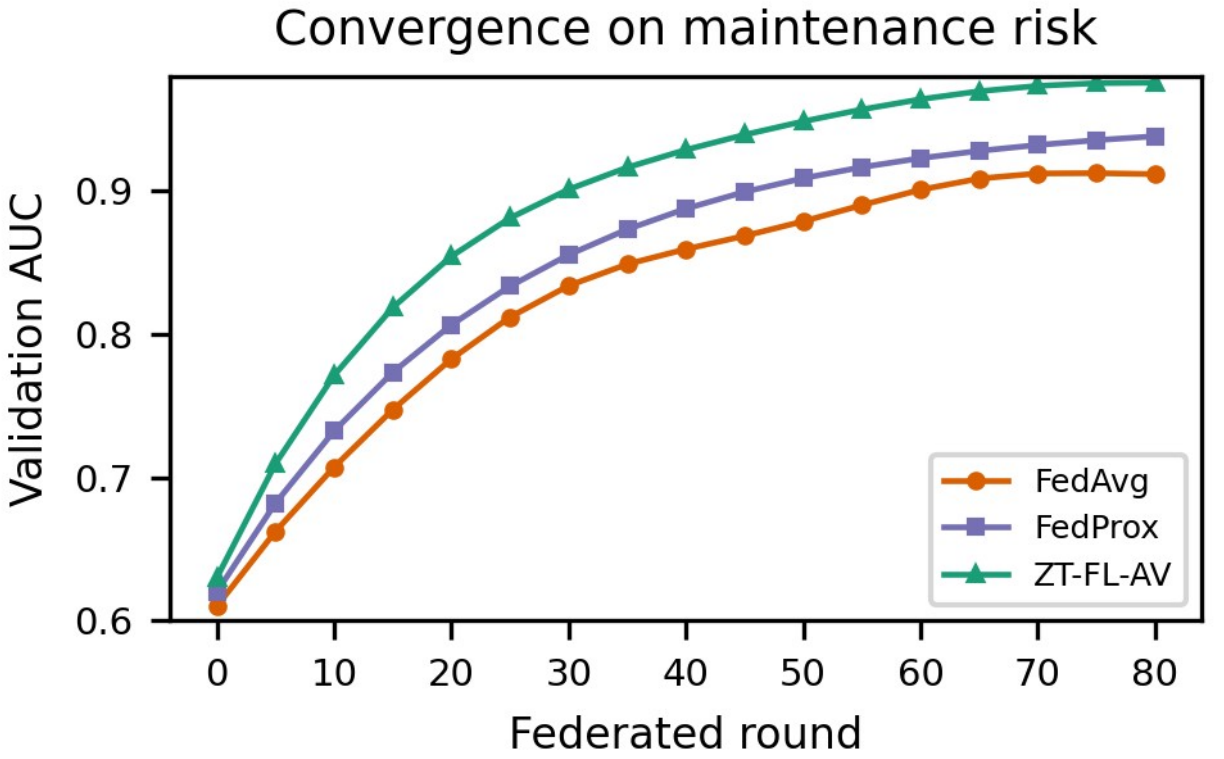

Fig. 3. Validation AUC over 80 federated rounds for maintenance-risk learning.
*Source/derivation: Series generated from the synthetic training log.*

Fig. 4 reports intrusion recall under model poisoning. At 0 percent malicious clients, all methods are near 0.90 recall. At 20 percent malicious clients, FedAvg drops to 0.612, FedProx to 0.681, and trimmed mean to 0.721. ZT-FL-CADE preserves 0.842 recall. The result demonstrates the intended behavior of the adversarial validator: the architecture sacrifices a small amount of benign-round simplicity in exchange for stronger resilience when training and OTA channels are attacked together. The result also shows why robust aggregation alone is insufficient. Trimmed mean improves over FedAvg but remains vulnerable to updates that are crafted to appear statistically plausible while damaging intrusion sensitivity.

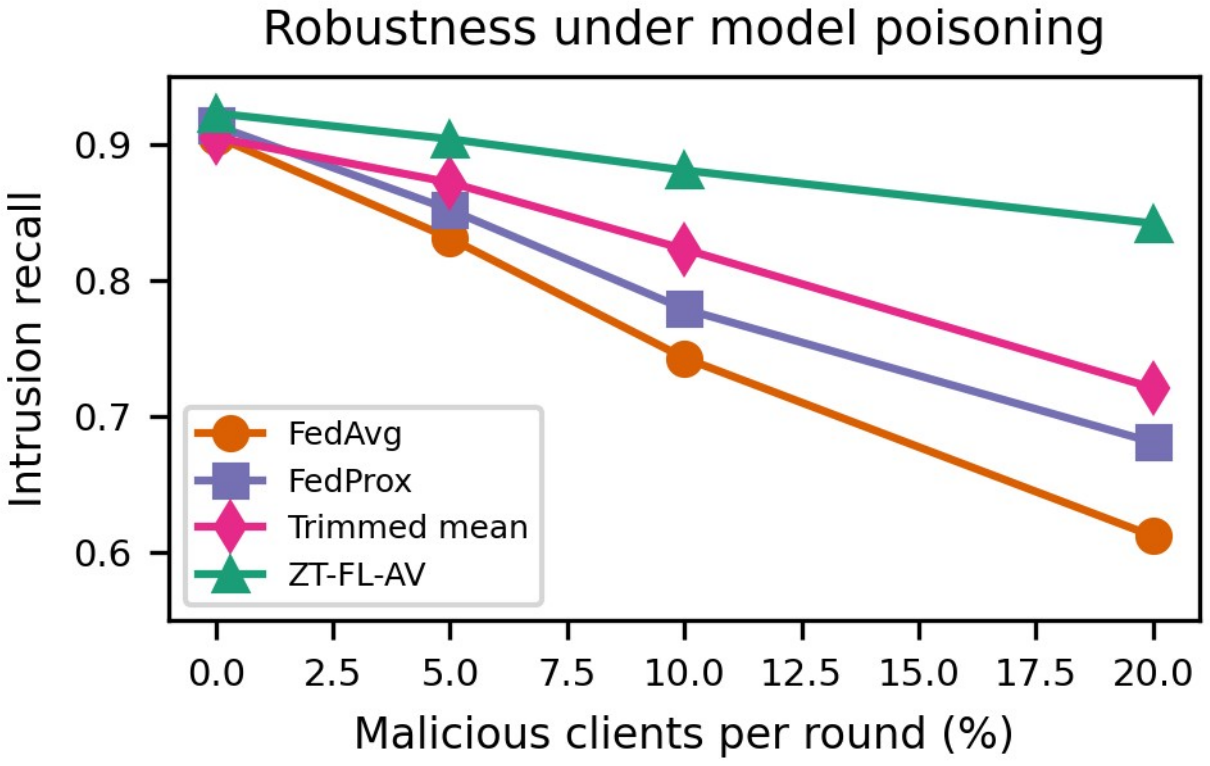

Fig. 4. Intrusion recall under increasing malicious-client participation.
*Source/derivation: Attack-intensity curves derived from the synthetic adversarial protocol.*

Access-control latency remains within the synthetic operational budget. The static PKI baseline has mean latency of 22 ms and 95th-percentile latency of 41 ms because it performs fewer checks. FedAvg plus rule-based trust has mean latency of 32 ms and 95th-percentile latency of 58 ms. ZT-FL-CADE has mean latency of 44 ms and 95th-percentile latency of 76 ms, as shown in Fig. 5. The added cost comes from model inference, posture feature computation, and ledger writing. A 2.6 percent false quarantine rate is observed on normal test update requests, below the 3.0 percent validation target. The challenge action absorbs many uncertain cases, reducing unnecessary quarantine while still interrupting risky update flows.

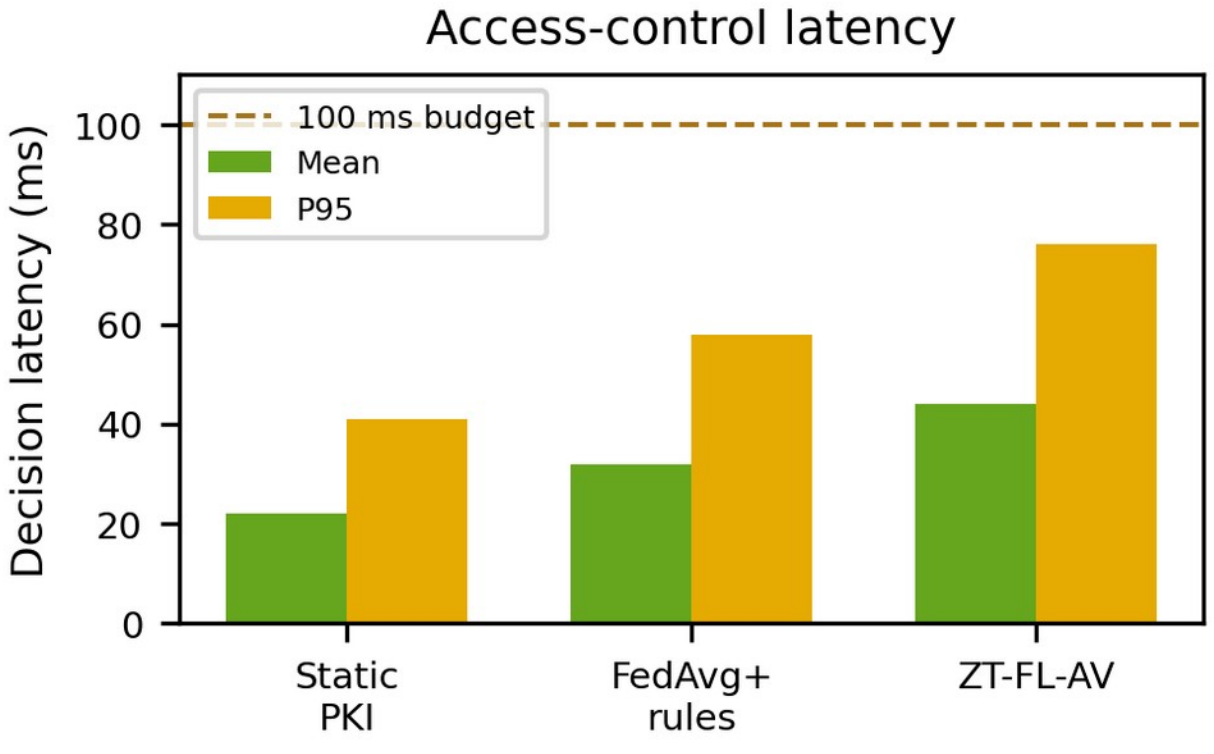

Fig. 5. Access-decision latency compared with the 100 ms synthetic budget.
*Source/derivation: Mean and 95th-percentile values computed from synthetic decision traces.*

The results also reveal tradeoffs. Differential privacy noise increases early-round variance and reduces the maximum validation AUC by roughly 0.006 compared with an otherwise identical non-private run. Secure aggregation increases communication expansion but avoids exposing individual updates. Adversarial validation can delay client contribution when a vendor experiences genuine distribution drift, such as a simulated heatwave affecting temperature and voltage patterns. For that reason, the framework includes a challenge path instead of only accept or reject. The challenge path can request re-attestation, additional validation windows, or human review before quarantine.

## VI. DISCUSSION

The synthetic results support a modest claim: in a controlled aftermarket-device simulation, combining zero-trust posture assessment with federated learning and adversarial validation improves the balance among maintenance prediction, intrusion detection, and OTA access control. The architecture is not presented as a substitute for secure boot, package signing, vulnerability management, supplier assessment, or regulatory cybersecurity processes. It is a complementary control plane that allows learned fleet behavior to influence device-level access decisions while preventing every local update from being trusted by default.

One practical implication is that trust scores should not be treated as opaque model outputs. In ZT-FL-CADE, the access action is accompanied by a short evidence record: model version, risk score, attestation age, signature retry count, drift score, selected policy rule, and validator decision. Such evidence can help engineering and security teams distinguish a suspicious OTA request from a normal device operating in a harsh environment. It can also support post-incident review when a quarantine action delays a repair or when an allowed update later appears problematic. This design aligns with AI risk-management expectations that governance and measurement should be embedded in the system lifecycle [26].

The framework also suggests a way to handle vendor heterogeneity without forcing a single data owner to centralize all telemetry. Vendors with small or specialized fleets can benefit from global failure patterns while retaining local records. The tradeoff is that FL governance becomes part of the security boundary. Client selection, update validation, privacy budgets, and model rollback procedures must be operated with the same discipline as OTA signing infrastructure. Prior ensemble and prescriptive analytics work supports the idea that multiple specialized learners can improve decision support when governance is explicit [27], [28]. Enterprise integration research adds a complementary lesson: observability and API governance should be designed early, because fragmented integration paths make incident response slower [29]. Process-automation governance work reaches a related conclusion for enterprise workflows that depend on coordinated controls across technical and organizational layers [30].

Several limitations remain. First, the dataset is synthetic. Feature ranges, class balances, and attack intensities were chosen to be plausible, not measured from deployed aftermarket fleets. Real telemetry will contain installation errors, seasonal patterns, regional network differences, diagnostic-tool variation, and missing records that may be harder than the simulation. Second, the adversarial model is limited to poisoning and update-abuse patterns. A stronger attacker could compromise attestation channels, mimic benign drift, or target the governance ledger. Third, the latency results assume an edge gateway with adequate compute resources. Very constrained devices may need model compression, coarser windows, or cloud-assisted validation.

Fourth, the evaluation does not prove compliance with ISO/SAE 21434 or UN regulations. It only identifies artifacts that could contribute to a compliance evidence package, such as traceable update decisions and logged model versions. Fifth, differential privacy and secure aggregation parameters are simulated at the architecture level. A production deployment would require formal privacy accounting, key-management design, dropout handling, and red-team testing. Finally, the policy thresholds were selected against validation data. Threshold governance must be revisited when the operating context changes, because an aggressive quarantine policy can reduce cyber risk while increasing maintenance disruption.

## VII. CONCLUSION AND FUTURE WORKS

This paper presented ZT-FL-CADE, a zero-trust federated-learning architecture for connected aftermarket devices. The framework links three decisions that are often handled separately: how devices participate in collaborative learning, how model updates are validated against adversarial behavior, and how OTA update requests are admitted, challenged, or quarantined. The architecture uses local feature extraction, privacy-aware federated training, adversarial update validation, device-posture scoring, and governance logging to support predictive maintenance and update-security workflows without raw-data pooling across vendors.

The synthetic evaluation used a single internally consistent dataset with 144,000 five-minute windows, 240 devices, 12 vendors, and 180 days of simulated operation. Within that setting, the proposed method improved maintenance-risk F1 from 0.837 for FedAvg to 0.883, improved intrusion F1 from 0.856 to 0.897, and preserved 0.842 intrusion recall under 20 percent malicious-client intensity. Access-decision latency remained below the 100 ms budget, with a 44 ms mean and a 76 ms 95th percentile. These results are promising as architectural evidence but should not be interpreted as field validation. They show that an integrated zero-trust and FL pipeline can be evaluated quantitatively, not that the exact numbers will transfer to production fleets.

Future work should proceed in four directions. The first direction is field-data validation with participating aftermarket vendors, repair networks, or fleet operators. Such validation should preserve privacy but include real installation histories, update failures, seasonal driving conditions, and technician workflows. The second direction is stronger adversarial evaluation. Poisoning attacks should be combined with compromised attestation, replayed update metadata, delayed client participation, and adaptive evasion against the validator. The third direction is formal governance. Privacy accounting, model-card style documentation, policy change control, and audit evidence should be attached to every global model release and OTA campaign. The fourth direction is implementation efficiency. Quantization, sparse updates, asynchronous aggregation, and tiered validation can reduce latency and bandwidth for lower-cost devices.

A broader research question concerns responsibility boundaries. Connected aftermarket devices often sit between original equipment manufacturers, part suppliers, distributors, repair shops, fleet owners, and consumers. A zero-trust FL system can produce technical evidence, but organizational agreements must define who can approve a quarantine, who reviews false positives, and who owns model rollback. Future prototypes should therefore include operational playbooks, not only algorithms. They should also test how human reviewers interpret the evidence record attached to allow, challenge, and quarantine actions.

The final contribution of this work is methodological. A synthetic dataset cannot replace deployment, but it can make architectural assumptions explicit. By fixing row counts, entity counts, feature ranges, class balances, splits, and evaluation metrics, the analysis avoids mixing incompatible numbers across sections. This discipline is important for emerging automotive AI security research, where claims about federated learning, zero trust, and adversarial robustness can otherwise become difficult to compare. Future studies can replace the synthetic generator with public or confidential field datasets while preserving the same evaluation template.